\documentclass[conference]{IEEEtran}
\usepackage{floatrow}
\usepackage{url}

\usepackage{cite}

\ifCLASSINFOpdf
\usepackage[pdftex]{graphicx}
\else
\fi
\usepackage[cmex10]{amsmath}
\usepackage[caption=false,font=footnotesize]{subfig}
\begin{document}
%
\title{Test Set Diameter:\\
Quantifying the Diversity of Sets of Test Cases}

\author{\IEEEauthorblockN{Robert Feldt and Simon Poulding}
\IEEEauthorblockA{Software Engineering Research Lab\\
Blekinge Institute of Technology\\
Karlskrona, Sweden\\
Email: robert.feldt@bth.se, simon.poulding@bth.se}
\and
\IEEEauthorblockN{David Clark and Shin Yoo}
\IEEEauthorblockA{Department of Computer Science\\
University College London\\
London, UK\\
Email: david.clark@ucl.ac.uk, shin.yoo@ucl.ac.uk}}


%


\maketitle

\begin{abstract}
A common and natural intuition among software testers is that test cases need to differ if a software system is to be tested properly and its quality ensured. 
Consequently, much research has gone into formulating distance measures for how test cases, their inputs and/or their outputs differ. 
However, common to these proposals is that they are data type specific and/or calculate the diversity only between pairs of test inputs, traces or outputs. 

We propose a new metric to measure the diversity of sets of tests: the test set diameter (TSDm). 
It extends our earlier, pairwise test diversity metrics based on recent advances in information theory regarding the calculation of the normalized compression distance (NCD) for multisets. 
An advantage is that TSDm can be applied regardless of data type and on any test-related information, not only the test inputs. 
A downside is the increased computational time compared to competing approaches. 

Our experiments on four different systems show that the test set diameter can help select test sets with higher structural and fault coverage than random selection even when only applied to test inputs. This can enable early test design and selection, prior to even having a software system to test, and complement other types of test automation and analysis.
We argue that this quantification of test set diversity creates a number of opportunities to better understand software quality and provides practical ways to increase it.

\end{abstract}


%
\IEEEpeerreviewmaketitle

\section{Introduction}

Testing is a critical activity to ensure that software has sufficient quality and reliability.
Since software projects often exceed the time and resources allocated to them, testing frequently has to make the most use of the least resources. This limits the number of test cases that can be selected even if automated techniques are used for test execution and/or when creating tests. Even if there is ample time to plan for testing and develop test cases, the tests will typically be re-executed frequently, within each iteration or prior to each release, which means that we need as small and as effective a set of tests as possible.
Thus, a fundamental problem in software testing is \emph{how to select a small set of test cases} that most efficiently tests a software system. In general, we also prefer test cases that are themselves small since a human will typically act as oracle and decide if the system's behavior is correct.


To achieve these goals a natural intution among testers is to select a set of small but diverse test cases. The idea is that test cases that differ are more likely to better cover the intended as well as the actual software behavior. This approach also has support in the research literature. For example, adaptive random testing~\cite{chen2010adaptive} only adds a new, randomly-generated test case if it has large distance to existing test cases. But Chen et al.~\cite{chen2010adaptive} also note that a number of testing methods such as Restricted Random Testing~\cite{chan2002restricted}, Antirandom testing~\cite{malaiya1995antirandom}, and Quasi-Random Testing~\cite{chen2007quasi} are all based on the same idea: `evenly spreading' test cases over the input domain. Critical to the success of these techniques is a genericly applicable diversity measure and Chen et al.\ go as far as saying that `We have come to realise that ``even spreading''
can be better described as a form of diversity'~\cite{chen2010adaptive}. They describe a distance calculation scheme based on the category-partition (partition testing) method, but it requires that the tester manually identifies categories and levels which can be varied.

Most approaches to quantifying diversity among test cases are specific to a certain type of data. It is common to assume that the data is numeric since there are a multitude of existing distance functions that can then be applied. One example is the approach of Bueno et al.~\cite{bueno2007improving} which selects test sets that maximize the sum of the distances from each test input to its nearest neighbor. In their empirical work they use the Euclidean distance which requires the inputs to be numerical vectors. More recently, Alshawan et al.~\cite{alshahwan2012augmenting} proposed to select test cases for web application testing based on unique system outputs. Their empirical study~\cite{alshahwan2014coverage} applied several uniqueness measures defined on the HTML output of five web applications. All of the uniqueness measures are binary in nature since they are based on comparing equality among parts of the HTML output.


Feldt et al.~\cite{Feldt2008:TestDiversity} proposed a test diversity metric with a stronger theoretical basis by using the concept of information distance from the field of Information Theory~\cite{bennett1998information}. Although information distance is based on Kolmogorov complexity~\cite{li2009introduction} which can only be approximated in practice, such approximations using modern compression algorithms have shown wide applicability~\cite{Cilibrasi05,cilibrasi2004algorithmic,cilibrasi2007google}. Apart from the theoretical motivation there are also practical benefits to their proposal: since information distance is universal and applies to any strings of data, it can be used regardless of data type investigated.
The benefit to developers of testing systems would be immediate. They need no longer develop specific metrics for each application domain or type of data. By basing diversity measurements on the normalized compression distance (NCD) automated testing systems can search for diverse tests in general~\cite{Feldt2008:TestDiversity}.


A fundamental gap in the proposal of Feldt et al.\ is how the resulting diversity metric is to be used. Since NCD is a pairwise metric it can calculate a distance matrix between all pairs of (here) test cases, but it is not clear how to leverage that information for selecting test cases. In their study, Feldt et al.\ used a general approach (also employed in the original papers that proposed the NCD metric) to cluster the studied objects based on the distances. The result of such an analysis is a tree that defines an approximate order among the test cases. But Feldt et al.\ did not investigate concrete ways for selecting test cases to form a small, diverse test set.

This paper applies more recent results in Information Theory that extends NCD to multisets~\cite{cohen2012normalized}. This gives a general approach to solve test selection problems that can work for any type of test input, output or other information collected about a test, such as its execution trace. 
Our conjecture is that such a theoretically motivated approach to test selection based on the diversity of a set can create more efficient test sets than heuristic methods such as the maximize-mean-diversity used by Bueno et al~\cite{bueno2007improving} and the maximize-min-diversity used in adaptive random testing~\cite{chen2010adaptive}. 
The NCD for multisets can calculate the diversity in the whole set in a theoretically well-founded manner instead of approximating it based on pairwise comparisons. And since the underlying distance metric itself can be used for any data type, the approach can also be applied for as yet unknown data types or in new application domains.



Section~\ref{sec:background_ncd} introduces NCD and its extension to multisets and Section~\ref{sec:test_set_diameter} defines test set diameter and a test selection procedure based on it. The empirical study in Section~\ref{sec:empirical_study} presents the research questions, software under test, experimental procedures, and results. Finally, we discuss the implications of our results in Section~\ref{sec:empirical_study} and place them in the context of related work in Section~\ref{sec:related_work}, before we conclude the paper in Section~\ref{sec:conclusions}.

\section{Normalized Compression Distance for Multisets}
\label{sec:background_ncd}

In this section we describe the concepts from Information Theory that are used in this work: the information distance, its practical approximation as the normalized compression distance (NCD), and the recent extension of NCD to multisets.

\subsection{Information Distance and its approximation}
\label{sec:ncd}

The Kolmogorov complexity of a string of symbols, $x$, is the length of the shortest program that outputs $x$~\cite{li2009introduction}. It is a measure of the information contained in $x$, and we denote it here as $K(x)$.  The conditional Kolmogorov complexity of $x$ given $y$, denoted $K(x|y)$ is the length of the shortest program that outputs $x$ given the input $y$.

Bennett et al.\ proposed \emph{information distance} as a similarity measure that is calculated using conditional Kolmogorov complexity~\cite{bennett1998information}.  For two strings, $x$ and $y$, the information distance is:
\begin{equation}
	\text{ID}(x,y) = \max\{K(x|y), K(y|x)\}
\end{equation}
In other words, the similarity between entities $x$ and $y$ is the length of the shortest program that converts $x$ to $y$, or the shortest program that converts $y$ to $x$, whichever is larger.  Bennett et al.\ show that information distance is a \emph{universal} measure of similarity in the sense that if any other admissable metric detects a similarity between $x$ and $y$, then so will the information distance.

The Kolmogorov complexity of a long string will be, in general, larger than that of a short string.  For this reason, the information distance between two long strings will be, in general, larger than that between two short strings.  In order to be able to compare similarities between pairs of strings across a range of sizes, Li et al.\ proposed the \emph{normalized information distance} (NID) that normalizes the information distance using the Kolmogorov complexity of the two strings \cite{li2004similarity}:
\begin{equation}
	\text{NID}(x,y) = \frac{\max\{K(x|y), K(y|x)\}}{\max\{K(x), K(y)\}}
\end{equation}
NID takes values in the interval $[0,1]$, where values closer to 0 indicate greater similarity.  This particular choice of normalization has the advantage that NID retains the characteristics of a metric, i.e.\ that (i) $\text{NID}(x,x)=0$ (identity axiom);  (ii) $\text{NID}(x,y) + \text{NID}(y,z) \ge \text{NID}(x,z)$ (triangle axiom); and, (iii) $\text{NID}(x,y)=\text{NID}(y,x) )$ (symmetry axiom).

In practice, is not generally feasible to determine the shortest program that outputs a given string, and thus the Kolmogorov complexities that are used to calculate NID.  Cilibrasi and Vit\'{a}yni describe a practical alternative: the normalized compression distance (NCD)~\cite{cilibrasi2005clustering}.

NCD is based on the observation that the degree to which a string can be compressed by real-world compression programs, such as gzip and bzip2, is a good approximation of its Kolmogorov complexity.  If $C(x)$ is the length of the string $x$ after compression by a chosen compression program, then NCD is given by:
\begin{equation}
	\text{NCD}(x,y) = \frac{C(xy) - \min\{C(x),C(y)\}}{\max\{C(x),C(y)\}}
\end{equation}
where $xy$ denotes the concatenation of $x$ and $y$.  NCD takes values in the range $[0,1+\epsilon]$ where $\epsilon$ is a small positive value that depends on the degree to which the compression program approximates the Kolmogorov complexity.


\subsection{Extending NCD to Multisets}
\label{sec:ncdm}

NCD is a pairwise metric: it measures the degree of similarity between two strings.  Of more interest in some situations is a notion of similarity between a multiset\footnote{We use the strict terminology of \emph{multiset} rather than \emph{set} since the collection may contain the same string more than once.} of strings.  This is the case in this paper where we are interested in the similarity (or conversely, diversity) of a set of test cases considered as a whole.

Recently Cohen and Vit\'{a}nyi have extended the notion of NCD to multisets for this purpose. For a multiset $X$, the NCD is calculated via an intermediate measure, $\text{NCD}_1$, as~\cite{cohen2012normalized}:
\begin{align}
 	\text{NCD}_1(X) &= \frac{C(X)-\min_{x \in X} \{C(x)\}}{\max_{x \in X}\{ C(X\setminus\{x\}) \}}
	\label{eqn:ncd1} \\
	\text{NCD}(X) &= \max\left\{\text{NCD}_1(X), \max_{Y \subset X} \{ \text{NCD}(Y) \} \right\}
	\label{eqn:ncdm}
\end{align}
By defining the NCD of a set of one string to be 0, the NCD of a multiset of two strings is identical to the pairwise NCD between the two strings.

Since the term $\max_{Y \subset X} \{ \text{NCD}(Y) \}$ requires the recursive evaluation of NCD for each proper subset of X, the calculation of NCD for multisets has time complexity $\mathcal{O}(2^N)$ where $N$ is the size of the multiset.  This is likely to make the calculation of NCD impractical for anything but the smallest multisets.

Cohen and Vit\'{a}nyi suggest instead an algorithm that has time complexity $\mathcal{O}(N^2)$ to approximate NCD for multisets~\cite{cohen2012normalized} using the intermediate measure $\text{NCD}_1$ defined in~\eqref{eqn:ncd1} above.
The algorithm starts from the multiset $Y_0 = X = \{x_1, x_2, \ldots, x_n \}$, and proceeds as:
\begin{enumerate}
	\item Find index $i$ that maximizes $C(Y_k \setminus\{x_i\})$.
	\item Let $Y_{k+1} = Y_k \setminus {x_i}$ .
	\item Repeat from step 1 until the subset contains only two strings.
	\item Calculate $\text{NCD}(X)$ as: $\max_{0 \le k \le n-2} \{\text{NCD}_1(Y_k)\}$.
\end{enumerate}


\section{Test Set Diameter}
\label{sec:test_set_diameter}

Information distance, and its practical realization as the normalized compression distance (NCD), has an appealing characteristic when used to assess the similarity (or diversity) of software tests: it is universal in the sense that it ``discovers all effective feature similarities or cognitive similarities between two objects'' \cite{bennett1998information}.  If information distance is used to measure diversity between inputs, it can therefore be applied to inputs of any data type and so may obviate the need for domain-specific distance metrics such as the Euclidean distance between numeric inputs used by Bueno et al.~\cite{bueno2007improving}.

The generic applicability and universality of the metric also permits a much wider range of information to be considered when determining the diversity of test cases.  This benefit was investigated in the earlier work of Feldt et al.\ on the application of NCD to software testing \cite{Feldt2008:TestDiversity}.  The authors proposed a variability model that identifies a wide number of aspects in which test cases may differ. These aspects include: the setup of the SUT prior to executing the test, the test inputs, the execution trace of the SUT, the SUT's output, and non-functional aspects of the execution such as the performance.  Feldt et al.\ proposed that any of these aspects, or combinations of them, carry information that may be used to assess the similarity of test cases, and an empirical investigation demonstrated the use of pairwise NCD for this purpose.

In this paper, we investigate how the extension of NCD for multisets described recently by Cohen and and Vit\'{a}nyi \cite{cohen2012normalized} enables similarity to be measured by a universal metric at the level of entire sets of test cases rather between pairs of test cases.

We introduce the term \emph{Test Set Diameter (TSDm)} for the NCD for multisets metric applied to a chosen aspect of the test cases in the test set.  Using the aspects identified in the variability model of Feldt et al., we may define a family of such metrics: Input TSDm where NCD is calculated on the multiset of test inputs, Output TSDm calculated from the multiset of outputs, Trace TSDm calculated from the multiset of execution traces etc.

In this paper we investigate the use of TSDm for selecting a diverse test set of a specified size from a larger pool of potential test cases.  For this purpose, we leverage the $NCD_1$ algorithm proposed by Cohen and Vit\'{a}nyi to approximate NCD for multisets that was described in section~\ref{sec:ncdm} above.  Since this algorithm uses the $\text{NCD}_1$ metric to approximate NCD, we refer to this procedure for selecting diverse test cases as $\text{TSDm}_1$.

The $\text{TSDm}_1$ procedure is provided with a pool of potential test cases, the size of the pool being as least as great as the size of the desired test set.  At each iteration of the algorithm a subset of the pool is created by removing a test case in order to maximise the $\text{NCD}_1$ metric as described in section~\ref{sec:ncdm}. Thus a sequence of test sets of decreasing size is created by the procedure, and the subset of the desired size in this sequence is then chosen as the diverse test set.

\section{Empirical study}
\label{sec:empirical_study}

We focus on evaluating the use of the test set diameter in selecting test cases based on input diversity, i.e. Input TSDm (I-TSDm). Our motivation is that this is a more ambitious and novel use case for automation in software testing. If possible it could allow the selection and thus design of test suites prior to even having started the implementation of the software system. As long as at least a partial specification and description of the interface of the software is present we could generate test inputs based on them and, via I-TSDm, select a small, diverse test set from them. Future work will investigate applications of TSDm to test execution traces (as in \cite{Feldt2008:TestDiversity}), outputs from the software under test (SUT), as well as to different combinations of test inputs, trace information, and outputs.



Below we describe the design of our experiments by detailing the five research questions, the four experimental subjects and the methods and measures common to all of the investigations. For each research question in turn, we then describe the specific methods used to answer it as well as the results.

\subsection{Research questions}

Since the test set diameter is a new test metric there are a number of both fundamental as well as practical experiments that are needed. The basic assumption implicit in the test heuristic `select diverse test cases' is that such test cases are more likely to cover diverse parts of the specification and thus of the software (since it implements the specification). At the most fundamental level we thus would expect that test sets with higher test set diameter have higher code coverage. If this is not the case it would seem very unlikely that we can exploit any TSDm metric to improve testing. The first research question we investigate is thus:

\textbf{RQ1 -- Correlation to code coverage}: Are higher levels of I-TSDm associated with higher levels of code coverage?

We do not expect any such correlation to be perfect, but there must exist at least some correlation that any automated method can then exploit. To investigate RQ1 we consider large numbers of sets of test inputs randomly sampled from an initial pool of randomly generated test inputs. We vary the size of the subsets to better understand the sensitivity to test set size. Since we have no reason to expect that either the TSDm values or the code coverage values are normally distributed, we use the Spearman rank correlation coefficient to study the association between TSDm and code coverage. This is a nonparametric dependence measure which is not sensitive to the underlying distributions of values~\cite{zar1998spearman}.

If we can find any correlation in RQ1, it is then natural to study if the $TSDm_1$ selection procedure described in section~\ref{sec:test_set_diameter} can be used in practice to select better test sets.

\textbf{RQ2 -- Structural coverage ability}: Do test sets selected based on I-TSDm lead to higher code coverage than randomly selected test sets?

We compare to random selection of test sets since this is the baseline technique available at a very early stage of software development. More sophisticated testing techniques would require an executable version of the software under test and are thus not comparable. To provide an upper limit for the level of coverage that could reasonably be expected, we compare the results to a post-hoc greedy algorithm (described in section~\ref{sec:empirical_study_measures} below) and normalize the attained coverage level in relation to its results. This way we can compare results between different softwares under test (SUT). For all three methods we study how quickly the coverage grows as the test set size increases.

During preliminary experimentation it was evident that the size of each test input is a key feature in determining diversity and thus of exploiting I-TSDm: the $TSDm_1$-based selection would include any long test inputs first in the test sets and then gradually include shorter test inputs. In retrospect this makes sense since, on average, the longer a test input is, the more room there is for it to be different from others. However, this means there is a risk that any effect seen when investigating RQ2 is due simply to the input size rather than to diversity \emph{within a given input size}.

\textbf{RQ3 -- Structural coverage ability w. size constraints}: Do test sets selected based on I-TSDm lead to higher code coverage than randomly selected test sets when we control for the size of test inputs?

Even though the test data generation technique we use~\cite{feldt2013godeltest} can be used to generate data with a specific size, we use here a simpler approach in order to avoid introducing unnecessary biases into the generated inputs: a large number of inputs are generated and only those close to a target size are selected.  We vary the target size and include generated inputs that are within 10\% of the target in the initial pool of test cases to select from. With this procedure we saw no correlation between size and selection order and so can study any effect of I-TSDm in better isolation.

\textbf{RQ4 -- Fault finding ability}: Do test sets selected based on I-TSDm lead to higher fault coverage than test sets based on random selection?

It is not enough to cover larger parts of a program; what we ultimately want is to find faults so that we can eliminate them. By studying a SUT for which there are seeded faults we can investigate any benefits in fault finding ability.


\textbf{RQ5: Selection time}: How does the time to execute the selection method scale as the size of the initial pool increase?

Even if the NCD for multisets metric that is the basis for our proposed I-TSDm metric is effective, it nevertheless scales as $\mathcal{O}(N^2)$ (see section~\ref{sec:ncdm}) which might limit it use in online test automation systems where a developer or tester is waiting for the system to finish.  For such systems, it may be more beneficial to use a random selection scheme if the time taken to measure I-TSDm could be better employed in executing more test cases.

\subsection{Software Subjects}
\label{sec:subjects}

\textbf{JEuclid} 3.1.9 is a Java library that renders images from MathML, an XML format for describing the presentation of mathematical equations \cite{JEuclid2015}. The library is exercised by an application that takes MathML as input and renders the equation it describes in SVG format as output. The inputs for this subject are generated based on the XML Schema specification of MathML 2.0; the reader is referred to \cite{Poulding2015GeneratingXML} for details.  Since rendering to SVG is only one of a number of capabilities JEuclid, we restrict our measurement of coverage to the JEuclid core rendering and document handling module rather than the entire library.  The core module has 11,556 non-comment, non-blank lines of code (SLOC).  

\textbf{ROME} 1.0 is a Java library for parsing and converting RSS and Atom: XML formats for syndication feeds \cite{ROME2015}. The library is exercised by an application that takes an Atom feed as input, parses the feed, and outputs the contents and structure as text.  The inputs are generated based on the XML Schema specification of Atom 1.0.  We measure coverage of the entire ROME library, which has 11,704 SLOC.  

\textbf{NanoXML} is a small Java library for parsing XML.  We use the version of NanoXML from the Software-artifact Infrastructure Repository maintained by the University of Nebraska-Lincoln \cite{SIR2015}.  The library is exercised by an application that is a variant of software provided by SIR for this purpose. The application takes XML as input, parses it, and outputs the contents and structure as text. These inputs are generated using the MathML generator that is also used for JEuclid.  We measure coverage of the entire NanoXML library, which has 1,630 SLOC. 

\textbf{Replace} is a C application from the Software-artifact Infrastructure Repository that performs pattern matching and substitution.  It takes three inputs: a string to be modified, a regular expression that defines matching text, and a string that replaces the matched text.  The regular expression is generated using a custom generator based on the specification of Replace; the string to be modified and replacement string are generated as random sequences of characaters of size between 0 and 32 characters.  The application has 538 SLOC.

\subsection{Measures}
\label{sec:empirical_study_measures}

All experiments were run on a MacBook Pro (Retina, Mid 2012) with a 2.7Ghz Intel Core i7 CPU and 16GB of DDR3 memory. The experimental framework as well as the TSDm and NCD related code were implemented in the \verb|Julia| programming language~\cite{JuliaLang2012}. For all experiments we used \verb|Julia| version 0.3.8 (2015-04-30). All experiments used 10 independent repetitions unless otherwise stated.

\subsubsection{Coverage}

JaCoCo 0.7.4 \cite{JaCoCo2015} was used to measure the structural coverage of the Java subjects: JEuclid, ROME, and NanoXML. JaCoCo collects coverage information at the bytecode level while the Java Virtual Machine is running and thus does not require the code to be instrumented. JaCoCo evaluates a number of structural coverage metrics: for this work the metric used was instruction coverage. 

For the Replace subject, coverage was assessed in terms of the detection of seeded faults in the application.  This assessment used 32 variants of application provided by the Software-artifact Infrastructure Repository: each variant contained one fault that had been manually seeded \footnote{For reasons related to the framework used to perform the experiments, 1 of the 32 variants was excluded and so considered undetected by any test case.}. A fault was considered to be detected by a test case if the output and/or return status from the variant differed from that of the unmodified application.

\subsubsection{Greedy Coverage Algorithm}

The greedy algorithm starts with an empty test set.  At each iteration, it considers all the test cases remaining in the pool and selects from these the test case that, if added to set, would improve the coverage the set achieves the most.  This test case is then removed from the pool and added to the test set.  This form of greedy algorithm is termed the `additional' approach in the context of regression testing \cite{ZhangH0RM13}.  While I-TSDm uses only test inputs themselves to select test cases, the greedy algorithm uses information obtained by dynamically executing the software under test using the test inputs.  For this reason we consider the coverage achieved by the greedy algorithm as the best that could be achieved using the given pool of test cases.


\subsection{Experiment 1: Correlation to code coverage}

In order to investigate the relationship between I-TSDm and code coverage we would like a wide a range as possible of test set diameters.  If we were to randomly sample subsets, it is likely we get mostly test sets with an average diameter and thus would be unable to assess the correlation over its full range of values.  Simply increasing the sample size is not feasible since it would take a very long time evaluate code coverage for all of them.

We thus opted for a stratified sampling of test subsets based on the I-TSDm selection sequence described in sections~\ref{sec:ncdm}.  The test inputs are considered in the order in which they are removed from the subsets $Y_k$ to form subset $Y_{k+1}$, and divided into 10 consecutive strata: strata 1 includes test inputs numbered 1 to 100, strata 2 include numbers 101 to 200, and so on.  Test sets are formed by randomly selecting a strata and then randomly sample inputs within the strata.  We repeat this procedure 100 times for each of the subset sizes 10, 25, and 50.

In table~\ref{table:correlation} we can show the Spearman rank correlation values for each of the three investigated SUTs and for each investigated subset size. In all cases we are able to reject the null hypothesis that there is no correlation between I-TSDm and instruction coverage at p-values less than $10^{-4}$, with the alternative hypothesis that the correlation is higher than 0.

\begin{table}[!t]
\renewcommand{\arraystretch}{1.3}
\caption{Spearman rank correlation values between I-TSDm and instruction coverage for three different SUTs and three different test set sizes.}
\label{table:correlation}
\centering
\begin{tabular}{|l||c|c|c|}
\hline

 & \multicolumn{3}{ |c| }{\textbf{Test Set Size}} \\
\hline

\textbf{SUT} & \textbf{10} & \textbf{25} & \textbf{50}\\
\hline

JEuclid & 0.59 & 0.67 & 0.52 \\
\hline

NanoXML & 0.50 & 0.40 & 0.26\\
\hline

ROME & 0.60 & 0.57 & 0.82\\
\hline

\end{tabular}
\end{table}

\begin{figure*}[!t]
\centering
\subfloat[JEuclid]{\includegraphics{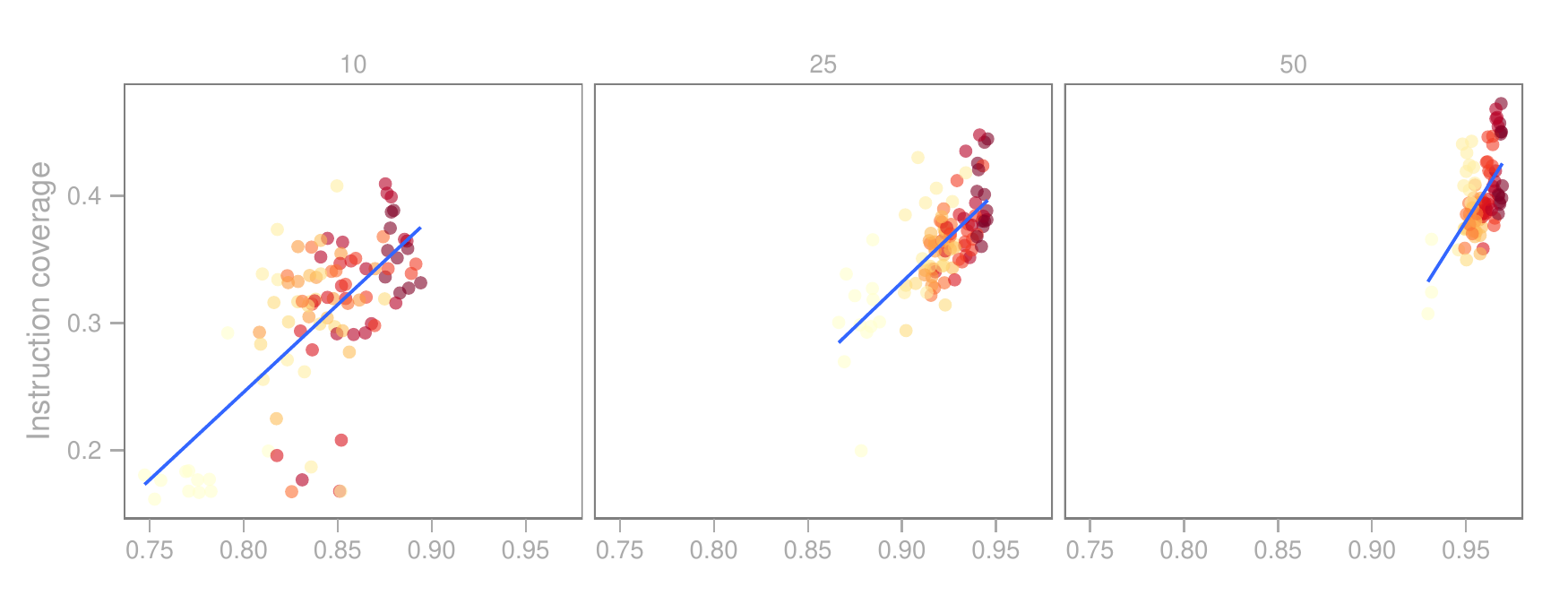}%
\label{fig1}}
\hfil
\subfloat[NanoXML]{\includegraphics{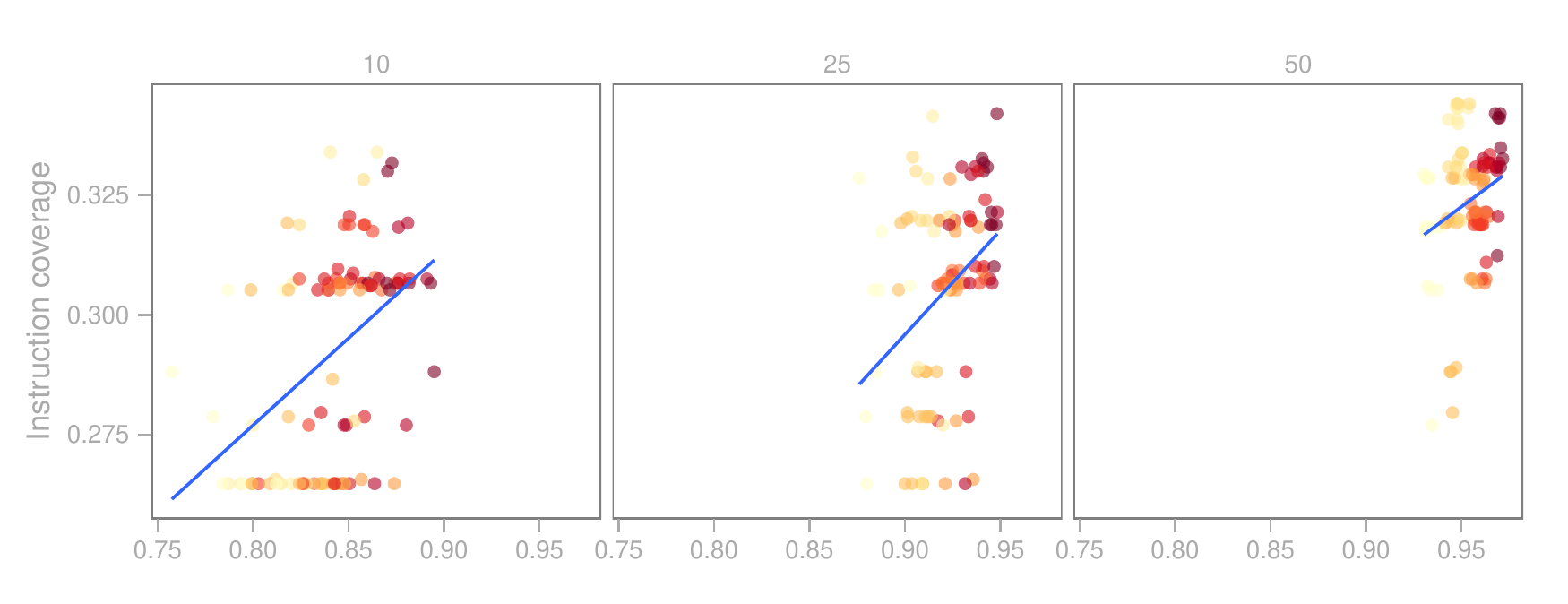}%
\label{fig2}}
\hfil
\subfloat[ROME]{\includegraphics{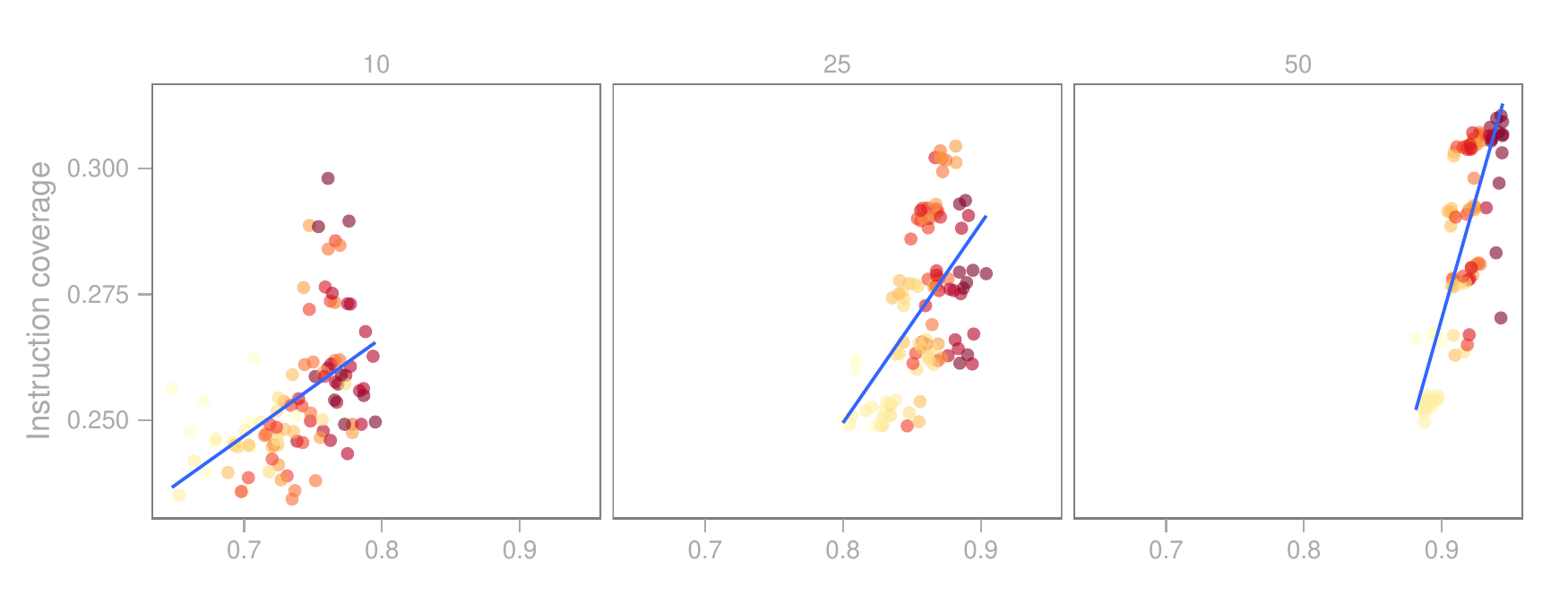}%
\label{fig3}}
\caption{Scatterplots showing the correlation between Input-TSDm (x axes) with instruction coverage (y axes) for test sets of three different sizes (10, 25, and 50 test inputs respectively) randomly sampled from 10 different strata. 
Test sets with darker colors are sampled from strata with test inputs that are selected by the I-TSDm selection procedure when the subset is the smallest.}
\label{fig:rome_corr_plots}
\end{figure*}

The correlation values are in the range 0.40 to 0.82 with one outlier (NanoXML at subset size 50) with a lower correlation value (0.26). Overall the correlation between I-TSDm and instruction coverage can be considered a moderate positive correlation. In Figure~\ref{fig:rome_corr_plots} the three sets for each subject are plotted next to each other using the same scales for the instruction coverage (y axis) and I-TSDm (x axis).  The pattern can be seen most clearly that as I-TSDm values increase on average as the test set size increase, and the same trend can be seen for the coverage values. For example, for ROME the average coverage is 0.254, 0.277, and 0.285 for test set sizes 10, 20, and 50 respectively.  Note also that the spread of TSDm values shrinks as the subset size increases and that there is in general more variation for the smaller subset sizes. The minimums and maximums for each subset size shows an increasing trend, although less so for NanoXML. Overall, NanoXML seems to be an outlier and its coverage values shows more of a banding effect.


\framebox(228,28)[c]{%
    \parbox{220\unitlength}{Test sets with higher input diameter (I-TSDm) on average have higher code coverage. 
    }
}

\subsection{Experiment 2: Structural coverage}


Since Experiment 1 demonstrated a correlation between I-TSDm and code coverage, we may reasonably expect that the TSDm ordering procedure can be used to select good test sets.  The results of Experiment 2 confirm that this is the case.

\begin{figure}[!t]
\centering
\includegraphics[width=3.1in]{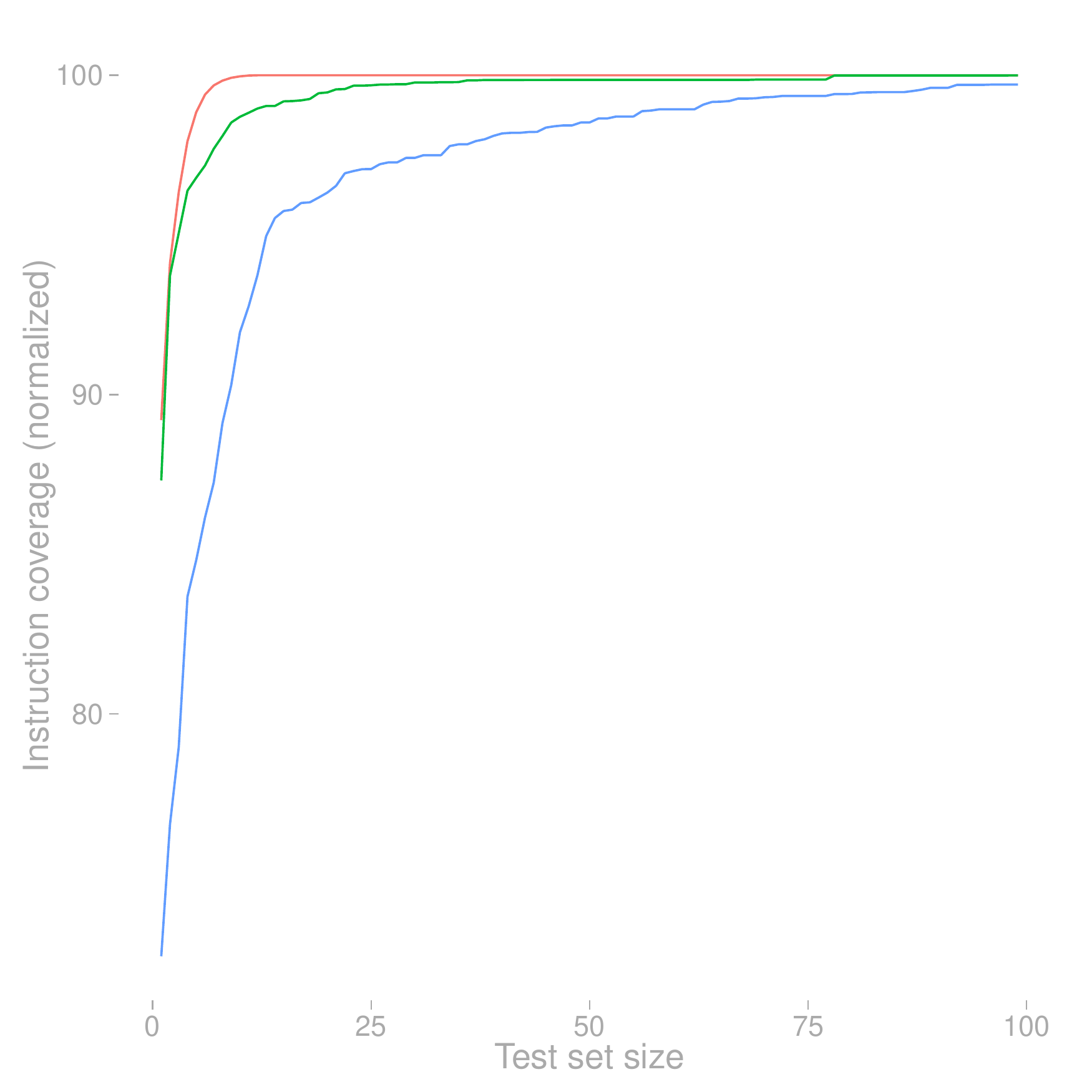}
\caption{Instruction coverage (normalized) of the ROME library against size for test sets selected using the greedy algorithm (red), $\text{I-TSDm}_1$ procedure (green), and random algorithm (blue) from an initial pool of 250 randomly-generated MathML inputs. The graphs are the average values over 10 repeated runs.}
\label{fig:rome_line_graph_no_size_constraint}
\end{figure}

Figure~\ref{fig:rome_line_graph_no_size_constraint} is a typical example of the coverage achieved by test sets selected using I-TSDm. The graph plots the instruction coverage of the ROME library against size for test sets selected using the greedy algorithm (red), $\text{I-TSDm}_1$ procedure (green), and random algorithm (blue) from an initial pool of 250 randomly-generated MathML inputs.  The results are averaged over 10 runs and normalized to the maximum coverage attained by the test sets derived using the greedy algorithm.

Table~\ref{table:avgsetsize_nosizeconstraint} shows the average size of the test sets that are needed to reach 90\%, 95\%, and 99\% of the maximum coverage using I-TSDm or random selection. On average, the random selection procedure require test sets that are 2 to 6 times larger than sets selected using I-TSDm. For example, test sets that achieve 95\% normalised coverage are 77\% (ROME), 80\% (NanoXML), and 83\% (JEuclid) smaller when selected using the $\text{I-TSDm}_1$ procedure than when selected randomly.


\framebox(228,36)[c]{%
    \parbox{220\unitlength}{Test sets selected for highest test input diameter lead to higher code coverage than randomly selected test sets.
    }
}

\begin{table}[!t]
\renewcommand{\arraystretch}{1.3}
\caption{Average test set size needed to reach 90\%, 95\%, and 99\% of the maximum instruction coverage reached by the greedy algorithm when selecting test inputs using the $\text{I-TSDm}_1$ procedure and the random algorithm from an initial pool of 250 inputs.}
\label{table:avgsetsize_nosizeconstraint}
\centering
\begin{tabular}{|l||c|c|c||c|c|c|}
\hline

 & \multicolumn{6}{ |c| }{\textbf{Avg. Test Set Size}} \\
\hline

 & \multicolumn{3}{ |c|| }{\textbf{I-TSDm}} & \multicolumn{3}{ |c| }{\textbf{Random}} \\
\hline

\textbf{SUT} & \textbf{90\%} & \textbf{95\%} & \textbf{99\%} & \textbf{90\%} & \textbf{95\%} & \textbf{99\%} \\
\hline

JEuclid & 3.6 & 6.8 & 49.3 & 20.2 & 39.2 & 100.8 \\
\hline

NanoXML & 3.5 & 8.7 & 75.3 & 13.9 & 43.1 & 183.8 \\
\hline

ROME & 1.4 & 3.0 & 12.5 & 8.8 & 13.1 & 58.5 \\
\hline

\end{tabular}
\end{table}

\begin{figure}[!t]
\centering
\includegraphics[width=3.1in]{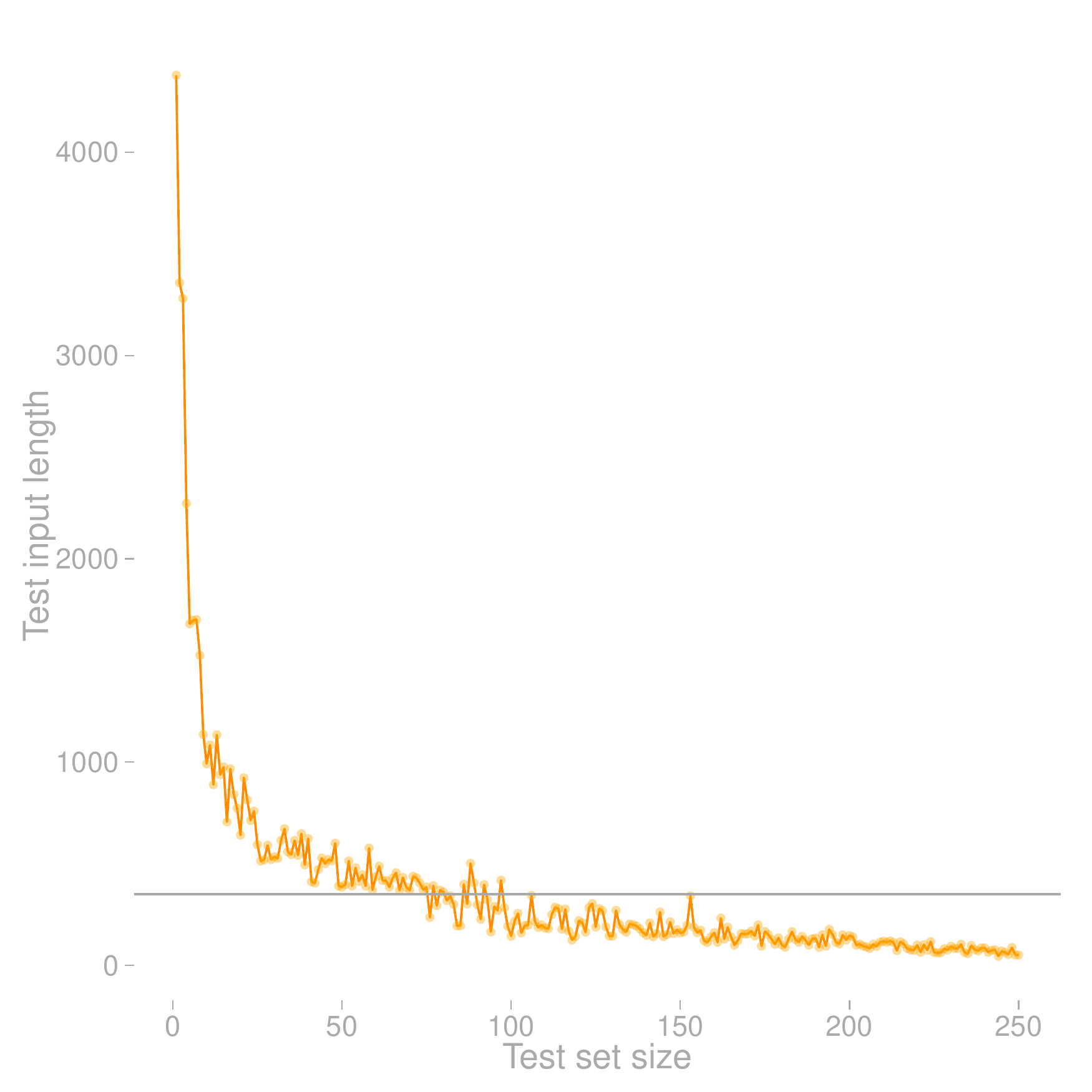}
\caption{Length of test inputs (for JEuclid) selected by the $\text{I-TSDm}_1$ procedure from an initial pool size of 250 plotted against the size of the test set in which they are first included, for one sequence of test sets. The mean value of 351.4 is indicated with the grey horizontal line.}
\label{fig:jeuclid_line_graph_no_size_constraint}
\end{figure}

However, if we look at the length of test inputs as selected by the $\text{I-TSDm}_1$ procedure we can see an almost perfect correlation between test case length and the size of the test set in which they are first included. Figure~\ref{fig:jeuclid_line_graph_no_size_constraint} plots the length of the test inputs for one such sequence of test sets (for JEuclid). The Spearman rank correlation was 0.96 in this case, 0.93 for ROME and 0.96 for NanoXML.

In hindsight this effect is rather obvious: the longer the input the more opportunity there will be for the input to exercise different parts of the software under test. This suggests that rather than using an elaborate test design strategy such as TSDm, a much simpler strategy could be used to select a few, large test inputs since they would appear likely to cover many aspects of the input space. This raises the question of whether diversity, and TSDm in particular, can offer any benefit beyond selection based on input length.

\subsection{Experiment 3: Structural coverage under a length constraint}

In this experiment, the benefit of I-TSDm is explored when the length of the test cases is restricted to a small range.  This corresponds to a requirement in practice to limit the length of test inputs: even when the test generation process is automated, the oracle is likely to be manual and thus the longer the test inputs the more difficult, and more costly, it will be to apply the oracle.  To evaluate if TSDm can help in this situation we re-ran Experiment 2 but limited the length of inputs in the initial pool to within 10\% of a target length using a generate-and-filter approach, i.e. random inputs were generated and inputs outside the allowed length interval were discarded.

\begin{figure}[!t]
\centering
\includegraphics[width=3.1in]{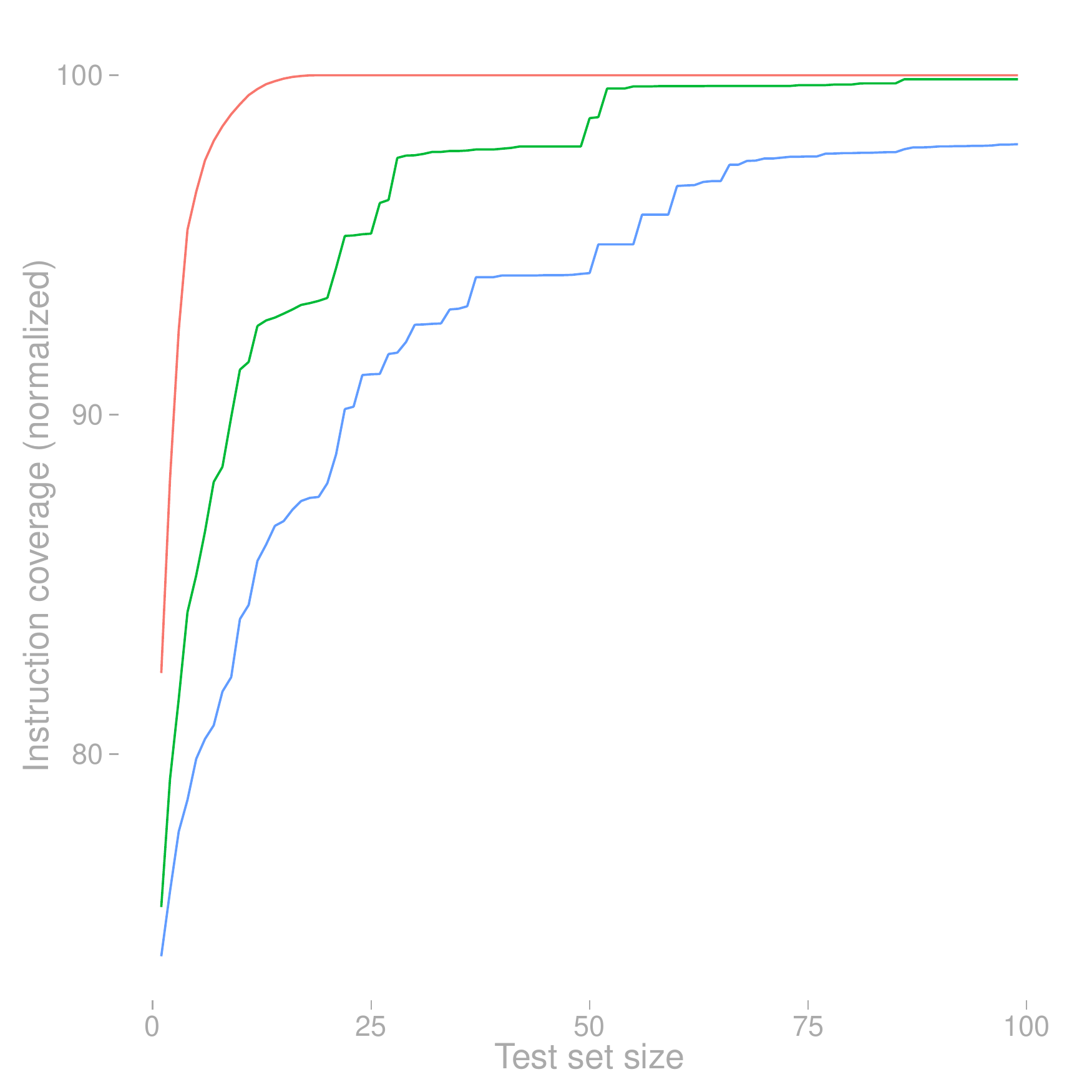}
\caption{Instruction coverage (normalized) of the ROME library against size for test sets selected using the greedy algorithm (red), $\text{I-TSDm}_1$ procedure (green), and random algorithm (blue) from an initial pool of 250 randomly-generated MathML inputs with lengths between 90 and 110 bytes. The graphs are the average values over 10 repeated runs.}
\label{fig:rome_line_graph_size_constraint100}
\end{figure}

Figure~\ref{fig:rome_line_graph_size_constraint100} shows the graphs for how instruction coverage (normalized) grows as the size of the test set increases.  This corresponds to Figure~\ref{fig:rome_line_graph_no_size_constraint} above but here the length of the inputs in the initial pool has been limited to range of 90-110 bytes.  We can see that the coverage of the test sets selected using I-TSDm is now less competitive compared to the greedy algorithm than when the input length was unconstrained. However, it still maintains a considerable advantage compared to randomly-selected test sets. We tested with several target lengths and the patterns persist until the target length is large enough and the advantage to select based on diversity diminishes. It seems that when each input is long enough to exercise a large part of the intended behavior we can pick any, i.e. random selection `catches up'.

Table~\ref{table:avgsetsize_sizeconstraint100} shows the average test set size to reach different levels of the maximum instruction coverage reached by any of these test sets.  We note that for a given coverage level, a test set selected using the $\text{I-TSDm}_1$ procedure is 2 to 9 times smaller than a randomly selected set. For example, test sets that achieve 95\% normalised coverage are 49\% (NanoXML), 57\% (ROME), and 70\% (JEuclid) smaller when selected using the $\text{I-TSDm}_1$ procedure than when selected randomly.

Comparing these results to Table~\ref{table:avgsetsize_nosizeconstraint}, it is clear that larger test sets are required than when the input length is unconstrained, and this occurs for \emph{both} I-TSDm and the random algorithm. For ROME a test set of approximately 22 inputs is required to reach the 95\% coverage using I-TSDm, while only 3 inputs are needed when the length is constrained.  However, we exercise caution in comparing these tables since the unnormalized coverage values reached differ for some of the SUTs. For NanoXML and ROME a similar coverage level is obtained with and without a length constraint---36.1\% and 32.8\% respectively; but for JEuclid a coverage of 46.8\% is reached when input length is constrained, and 59.8\% when it is unconstrained. 

\begin{table}[!t]
\renewcommand{\arraystretch}{1.3}
\caption{Average test set size needed to reach 90\%, 95\%, and 99\% of the maximum instruction coverage reached by the greedy algorithm when selecting test inputs using the $\text{I-TSDm}_1$ procedure and the random algorithm from an initial pool of 250 inputs with lengths between 90 and 110 bytes.}
\label{table:avgsetsize_sizeconstraint100}
\centering
\begin{tabular}{|l||c|c|c||c|c|c|}
\hline

 & \multicolumn{6}{ |c| }{\textbf{Avg. Test Set Size}} \\
\hline

 & \multicolumn{3}{ |c|| }{\textbf{I-TSDm}} & \multicolumn{3}{ |c| }{\textbf{Random}} \\
\hline

\textbf{SUT} & \textbf{90\%} & \textbf{95\%} & \textbf{99\%} & \textbf{90\%} & \textbf{95\%} & \textbf{99\%} \\
\hline

JEuclid & 29.9 & 40.9 & 90.3 & 82.2 & 135.3 & 217.3 \\
\hline

NanoXML & 1.9 & 19.4 & 75.1 & 18.7 & 38.2 & 207.2 \\
\hline

ROME & 9.1 & 21.7 & 51.3 & 21.9 & 51.0 & 129.0 \\
\hline

\end{tabular}
\end{table}

\framebox(228,36)[c]{%
    \parbox{220\unitlength}{Test set diameter can lead to higher code coverage even if we control for the size of test inputs; test diversity is more than simply the input length.
    }
}

Given the importance that the length of a test case and a whole test set has automated testing techniques should preferably be compared by their coverage ability per bit (or byte) of test suite. 

\subsection{Experiment 4: Fault-finding ability}

For effective testing, structural coverage is not enough; ultimately the objective is to find faults. In this experiment we used the variants of Replace SUT with seeded faults to compare the fault-finding ability of the different procedures for selecting test inputs.
As discussed in section~\ref{sec:subjects}, this SUT takes three inputs, but the regular expression is by the far the most important in determing the execution path.  We therefore consider, and control for, only the length of this input.

\begin{figure}[!t]
\centering
\includegraphics[width=3.1in]{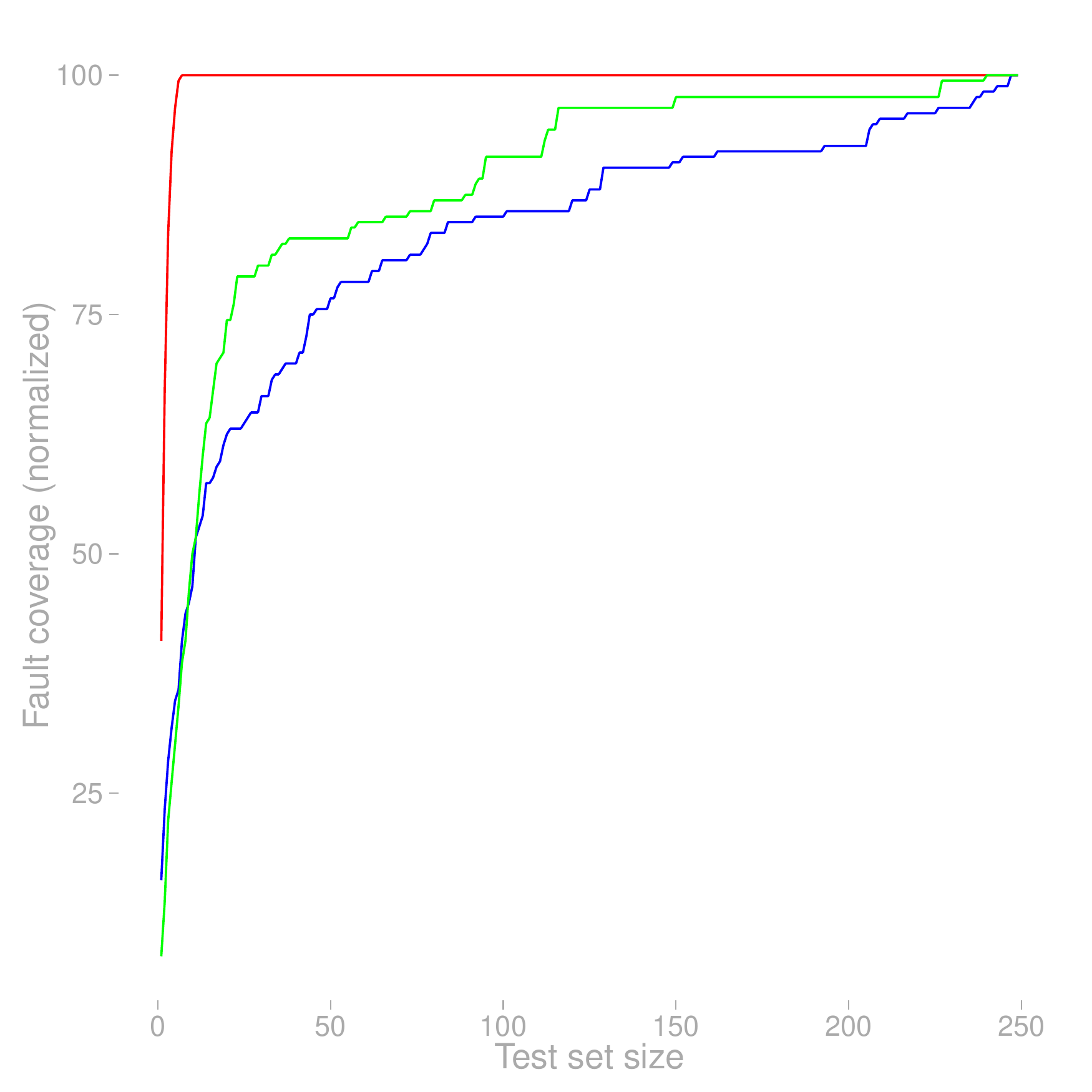}
\caption{Fault coverage (normalized) of the Replace SUT against size for test sets selected using the greedy algorithm (red), $\text{I-TSDm}_1$ procedure (green), and random algorithm (blue) from an initial pool of 250 randomly-generated inputs with the length of the regular expression between 9 and 11 bytes. The graphs are the average values over 10 repeated runs.}
\label{fig:sirreplace_constrained10_and_20}
\end{figure}

Figure~\ref{fig:sirreplace_constrained10_and_20} shows that I-TSDm selects test sets with higher fault coverage than randomly selected ones, on average, and the greedy algorithm has a fault-finding ability that is substantially better than both. Test sets that achieve 95\% normalised fault coverage are 45\% smaller when selected using the $\text{I-TSDm}_1$ procedure than when selected randomly.

\framebox(228,28)[c]{%
    \parbox{220\unitlength}{Test sets with larger test set diameter (I-TSDm) may have better fault-finding ability. 
    }
}

Of course, additional experiments that apply I-TSDm to the selection of test sets for a wider range of SUTs is necessary to be confident of the generality of this result.  

\subsection{Experiment 5: Test set selection time}

By collecting the run times of the I-TSDm test selection procedure during the experiments above, we are able to model its performance.  Since we expected a $\mathcal{O}(N^2)$ overall scaling (see section~\ref{sec:background_ncd}), we investigated a few different models that included this term. We found that a model of the form $S_{avg} * N^2$ where $S_{avg}$ is the average length of the strings being selected and $N$ is the number of elements in the initial pool explains the observed times extremely well. This is natural; we expect a linear scaling in the average length of the elements being compressed, as well as the $\mathcal{O}(N^2)$ scaling from the NCD1 reduction procedure. The $R^2$ goodness-of-fit had a value of over 0.99 when we fit this model with ordinary least squares regression.

\framebox(228,36)[c]{%
    \parbox{220\unitlength}{The TSDm test selection procedure scales quadratically in the size of the initial pool of tests to select from, and linearly with the average length of the tests.
    }
}

In practice this means that ~40-60 minutes of compute time were required on the machine we experimented on for an initial pool of 1000 test inputs with an average size of a couple of hundred bytes. Future work should investigate ways to speed up the process; possibly one can parallelize the calculations and/or exploit approximations based on initial filtering using pairwise NCD calculations.

\section{Discussion}

We have proposed the test set diameter (TSDm) as a family of diversity metrics with general applicability for software test generation and analysis. Our empirical study applied TSDm to the selection of test sets from a pool of randomly-generated test inputs and confirmed that it has a moderate to high correlation with structural code coverage. An important result from the study is the superiority of TSDm selection procedure compared to selecting test sets at random from the pool. For three Java software systems that take XML-based inputs, TSDm selects test sets that are 2 to 9 times smaller than randomly-selected sets that reach the same, high level of coverage. And on a software function implemented in C using a pool of inputs generated by a naively-implemented generator, TSDm demonstrated improved fault coverage compared to random selection. 
However, calculating TSDm is not cheap; even its approximation scales quadratically with the size of the pool from which test cases are selected.

Our results have implications for both theory and practice. There is a general lack of theories in Software Engineering~\cite{sjoberg2008building}, but as already noted by Chen et al~\cite{sjoberg2008building} when discussing the merits of adaptive random testing, diversity is a common thread to a large number of results in software testing. By taking a theoretically well-founded perspective of diversity based on Information Theory~\cite{li2009introduction,bennett1998information}, we can formalize what diversity means in software testing and explore if there are limiting laws. As we have argued elsewhere~\cite{clarke2015nier} there are a number of connections from Information Theory to software testing but diversity quantification is an important one and should be explored further.

For practice, there are numerous implications of our results. It is clear that TSDm can be used to select test sets that are more effective than randomly selected ones. Even if this will not always hold true there seems to be little risk in applying I-TSDm since there are so few alternative techniques that apply at early software development phases. Even if testers will continue to manually select test cases they can then use TSDm to search for additional test cases that most increase the overall test set diameter~\cite{Afzal2009,feldt2013godeltest}. This process could even be interactive: the tester is iteratively presented with a selection of highly diverse test cases to choose from~\cite{Feldt1999GPExplorative,Marculescu2015InteractiveSBST}.

An intriguing practical implication of our results is that we can use TSDm to analyse existing test suites. For example, ordering manually created test cases in a test set by TSDm, visualizing the results and discussing them with the testers might lead to insights into the relative importance of test cases in a similar way to what have been achieved for historical test outcomes~\cite{feldt2013supporting}. This can also help guide reductions of the test set as well as evaluation of newly proposed test cases that would most increase test set diversity.

Apart from the quadratic scaling of the test selection procedure time, there are a number of other limitations to our approach. One disadvantage to using NCD is that it is unclear what is the basis for a difference between two test cases. Test sets are not solely used for finding faults or ensuring coverage, but are also important elements in arguments for the quality, reliability and safety of a software system. For this use case there would be a benefit in a method that could not only select test cases but could explain why it has done a certain selection. Creating such a system based on NCD seems harder than creating one based on data type specific metrics. For example, if the diversity metric used is the size or structural depth of the output this can presumably be more readily understood by a human than if the explanation was based on describing the non-co-compressability of inputs. We can envision that metrics with higher explanatory power could be used after a TSDm based selection, but it seems clear that the inner workings of NCD are more opaque than data- and domain-specific metrics to which a user can relate.

Another limitation is that we do not yet understand the effect that the choice of compression algorithm has on NCD metrics. In informal investigations we have found TSDm results to be robust to variations in the compression algorithm used. However, many compression algorithms have some idiosyncratic design or implementation choices that might make them more or less suitable for the NCD calculation. For example, we tried to switch the default Zlib compressor for some of the faster ones in the Blosc library~\cite{Blosc2015}. However, they do not compress at all if the input string is less than 128 bytes. Such discontinuities might negatively impact the value from applying NCD based metrics. Preliminary investigations indicate that this is less of a problem for NCD for multisets that for pairwise NCD calculations since the former concatenates several strings and thus, in general, leads to longer strings.

The generality of NCD for multisets might make it attractive also to analyze and exploit non test-related software information in testing. For example, the text that can be found in software artefacts such as requirements, specifications, design documents, and source code might be used to find both test cases that are close to or different from such information. However, we note that the base NCD measure might not be applicable for things that are named rather than self-describing. Cilibrasi and Vit\'{a}nyi have proposed to instead use the Normalized Web Distance for such information~\cite{cilibrasi2009normalized}. Exploring this metric for software testing purposes remains an important area of future work.

An alternative method against which to compare I-TSDm test set selection would be manual selection, i.e.\ test sets created by humans. For two of the SUTs we have studied the Software-artifact Infrastructure Repository (SIR) provides test sets, but the test strategy used to create the test cases is not always clear, and we could not reliably use them for this purpose.
For this study we thus consider maximum coverage level attained by the greedy selection algorithm as a good approximation of what a human tester could achieve. Future research should nevertheless investigate the comparison to manual selection in more detail.

\section{Related work}
\label{sec:related_work}



The notion that more diverse test sets are beneficial for software testing has been around for some time; what has varied in the literature is the definition of the distance on which the diversity is defined. In a very limited sense, diversity can be interpreted as trying not to use the same subset of test cases repeatedly. Kim and Porter implemented this idea as the history-based test prioritization, in which the chance of selecting a test case is proportional to the elapsed time since its last execution~\cite{Kim2002HistoryRT}. Yoo et al.\ argued that repeated use of the same subset can `wear out' the selected test cases, and recommended using different subsets at each test iteration~\cite{Yoo:2009it}. While these works can be considered to concern test diversity, they are limited to making the best use of a given set of tests and cannot help test engineer to compare two sets of tests.

Test execution traces have been used as the basis of a richer diversity measure. Leon and Podgurski~\cite{Leon:2003vn} used the proportional binary distance (a variation of Euclidean distance for binary strings) to cluster test coverage vectors, and concluded that the more diverse the chosen subset of tests is, the higher the fault detection capability is. Nikolik has incorporated both control and data flow information to the definition of test diversity~\cite{Nikolik06testdiversity}. Ciupa et al.\ introduced the \emph{object distance} for object oriented programming languages by defining the distance between two obejct instances as the weighted sum of the distances between primitive members~\cite{Ciupa:2006rt}. Bueno et al~\cite{bueno2007improving} introduced a pairwise test set diversity measure based on Euclidean distance, and applied a metaheuristic search in the context of adaptive random testing. In a later study the authors mention that other distance measures could be used but the empirical work investigates numerical vectors only~\cite{bueno2014diversity}. The common thread in these existing works is the underlying use of Euclidean, Hamming, or Levenshtein distance measures, which either limits the applicability of the technique to numerical data, or looses high-level semantics. In contrast, the current paper uses a universal distance measure, NCD, which is free from the dependence on numerical vectors and proven to handle high-level semantics in the data.

Obtaining test execution traces requires instrumentation of the SUT. Alsahwan and Harman focused instead solely on the output of the SUT~\cite{alshahwan2012augmenting,alshahwan2014coverage} and showed that the test set with more diverse output can produce correspondingly higher fault detection capability and structural coverage. While Alsahwan and Harman's work focused on the diversity in the observed output, the current paper focuses on the diversity in the test \emph{input}, thereby allowing the analysis to take place even before test execution or system implementation. Since multiset NCD can be applied also to outputs a more fundamental difference is that Alshawan and Harman's output uniqueness measures are essentially binary since they only judge if two output are different or not. In contrast, the NCD metric used here (and previously applied in testing in~\cite{Feldt2008:TestDiversity}) can also quantify \emph{how} different, for example, outputs are. This should be important for software systems with large output spaces where most inputs are likely to return unique outputs or, in general, since it gives a more graded response than binary. Alshawan and Harman's measures instead filter the outputs before judging uniqueness and can thus consider uniqueness for different aspects. However, filters can also be used prior to applying NCD as noted by~\cite{Feldt2008:TestDiversity}. Future work should investigate the relative benefits of filters and/or binary uniqueness versus continuous measures such as NCD, both for inputs and for outputs.

Feldt et al.~\cite{Feldt2008:TestDiversity} is probably the closest existing work to the current paper. This work applied NCD to test cases in a pairwise manner to analyse the similarity between tests. Consequently, it could not help the test engineer to compare two different sets of test cases. The current paper extends the pairwise NCD to multiset NCD, enabling such comparisons.

\section{Conclusions}
\label{sec:conclusions}

Even though it has been long-argued that test sets with more diverse test cases are better, this notion has remained fuzzy and has been hard to apply in practice. Previous research has either focused on measuring distance only between pairs of test cases, or has been limited to specific data types. In this paper, we propose a metric called the test set diameter (TSDm) that measures the diversity of the test set as a whole. It has a formal basis in Kolmogorov complexity and applies to any data type and source of test related information. By approximating its calculation using modern compression algorithms it can be put to practical use for test selection and analysis.

Our empirical work evaluated the TSDm applied to test inputs. This is one of the more ambitious and difficult tasks in automated testing; one in which we need not even have started to implement the software we are going to test. Our results show that the input TSDm measure shows moderate to high positive correlation to instruction coverage for three open-source Java systems. A test selection procedure based on TSDm can approach a post-hoc greedy test selection procedure that needs actual coverage information from all test inputs in an initial pool. However, TSDm's success appeared to be in part achieved by selecting large test inputs early which put the value of diversity in question. But by controlling for the size of inputs to select from we show that TSDm still gives a considerable advantage compared to random selection. Finally, we applied the input TSDm selection procedure to a software implemented in C which has seeded faults. There is a positive effect on fault coverage: test sets that reach the same fault coverage level are 27 to 45\% smaller than those selected at random. 

We argue that being able to quantify and rank test sets based on their diversity is an important concept for software quality in general. However, diversity quantification is not enough; based on our empirical study we conjecture that a number of interrelated factors together determine the effectiveness of a test set including the diversity of the test data generator and the size and diversity in the initial pool selected from. Future work should evaluate them in detail and quantify their effect. We propose that this can lead both to a more fundamental understanding of test quality as well as practical techniques that improve it.

\section{Acknowledgments}
This work was partly funded by The Knowledge Foundation (KKS) through the project 20130085 Testing of Critical System Characteristics (TOCSYC).

\bibliographystyle{IEEEtran}
\bibliography{feldt_et_al_2015_test_set_diversity}

\end{document}